\documentclass{aastex631}

\shorttitle{V694 Mon is now burning nuclearly}
\shortauthors{Ulisse Munari}

\begin{document}

\title{The symbiotic star V694 Mon has transitioned from the
            accreting-only state\\ to the steady burning phase}

\email{ulisse.munari@inaf.it}
\author[0000-0001-6805-9664]{Ulisse Munari}
\affiliation{INAF Astronomical Observatory of Padova, 36012 Asiago (VI), Italy}

\begin{abstract}

{I propose that the drastic photometric and spectroscopic changes affecting
the symbiotic star V694 Mon since 2018, are due to its transition from the
accreting-only state to steady hydrogen-burning on the surface of the white
dwarf, closely mimicking the pattern followed by V4368 Sgr.  The phase of
peak optical brightness and weakest emission lines has probably been reached
in early 2024.  The high-velocity absorptions powered by jet-ejection and
the wild flickering, which dominated the century-long quiescence, should not
reappear as long has nuclear burning will hold (time-scale of decades).  The
3,500~L$_\odot$ burning luminosity suggests a 0.60~M$_\odot$ mass for the
WD.}

\end{abstract}

\keywords{symbiotic stars --- novae --- nuclear burning --- accretion}

\section{Early history of V694 Mon} \label{sec:intro}

V694 Mon (= MWC 560 = SS73 5) is a symbiotic star (SySt), in which an M4III cool
giant is orbited by a white dwarf (WD) surrounded by a massive accretion
disk \citep{2020MNRAS.492.3107L}.  It was first noted by
\citet{1943ApJ....98..153M} and \citet{1973ApJ...185..899S} as showing
Balmer emission lines flaked by wide and deep blue-shifted absorptions. 
\citet{1984BAAS...16..516B} reported about terminal velocities up to $-$3000
km/s and noted the persistent presence of photometric flickering, with an
amplitude of $\sim$0.2 mag and time-scales of a few minutes.  According to
the historic lightcurve compiled by \citet{1991IBVS.3563....1L}, V694 Mon
was highly variable around a mean $B$=12.0 mag brightness during 1925-1990. 
Suddenly in 1990, V694~Mon underwent a $\Delta$$B$$\sim$2mag surge in
brightness, never returned to prior fainter levels, and continued the
incessant flickering and the wild and chaotic variability around some
persisting periodicities \citep{2016NewA...49...43M}.

The 1990 surge in brightness coincided with an exacerbation of the velocity
and complexity of the blue-shifted absorptions, that reached a record
$-$6500 km/s \citep{1990IAUC.4987....1S}.  \citet{1990Natur.346..637T}
attributed the wildly variable absorptions to discrete jet-like ejections
with a relatively high degree of collimation and with the direction of the
ejection near the line of sight, with the binary system being seen close to
face-on conditions.  V694 Mon then remained self-similar from 1990 to 2018,
with typical optical spectra and line profiles for 2016 being shown in the
top and right panels of Fig.~\ref{fig1}: a hot continuum from the accretion
disk dominates in the blue, the molecular bands of the M giant emerge at
reddest wavelengths, and strong FeII and Balmer emission lines are flanked
by high-velocity and blue-shifted absorptions, while large and wild
flickering has been persistently affecting photometric observations
\citep{1996A&AS..116....1T,2001MNRAS.326..553S}.

\section{The drastic change which started in 2018} \label{sec:2018}

The appearance radically changed in 2018, when V694~Mon entered a steep rise
in brightness accompanied by the disappearance of both flickering and
high-velocity absorptions \citep{2018ATel12227....1G}.  Flickering and
high-velocity absorptions never reappeared afterward
\citep{2019ATel13236....1Z,2021ATel15066....1M}, while the once-strong
emission lines progressively weakened toward near-disappearance.

\begin{figure*}
\includegraphics[angle=270,width=18cm]{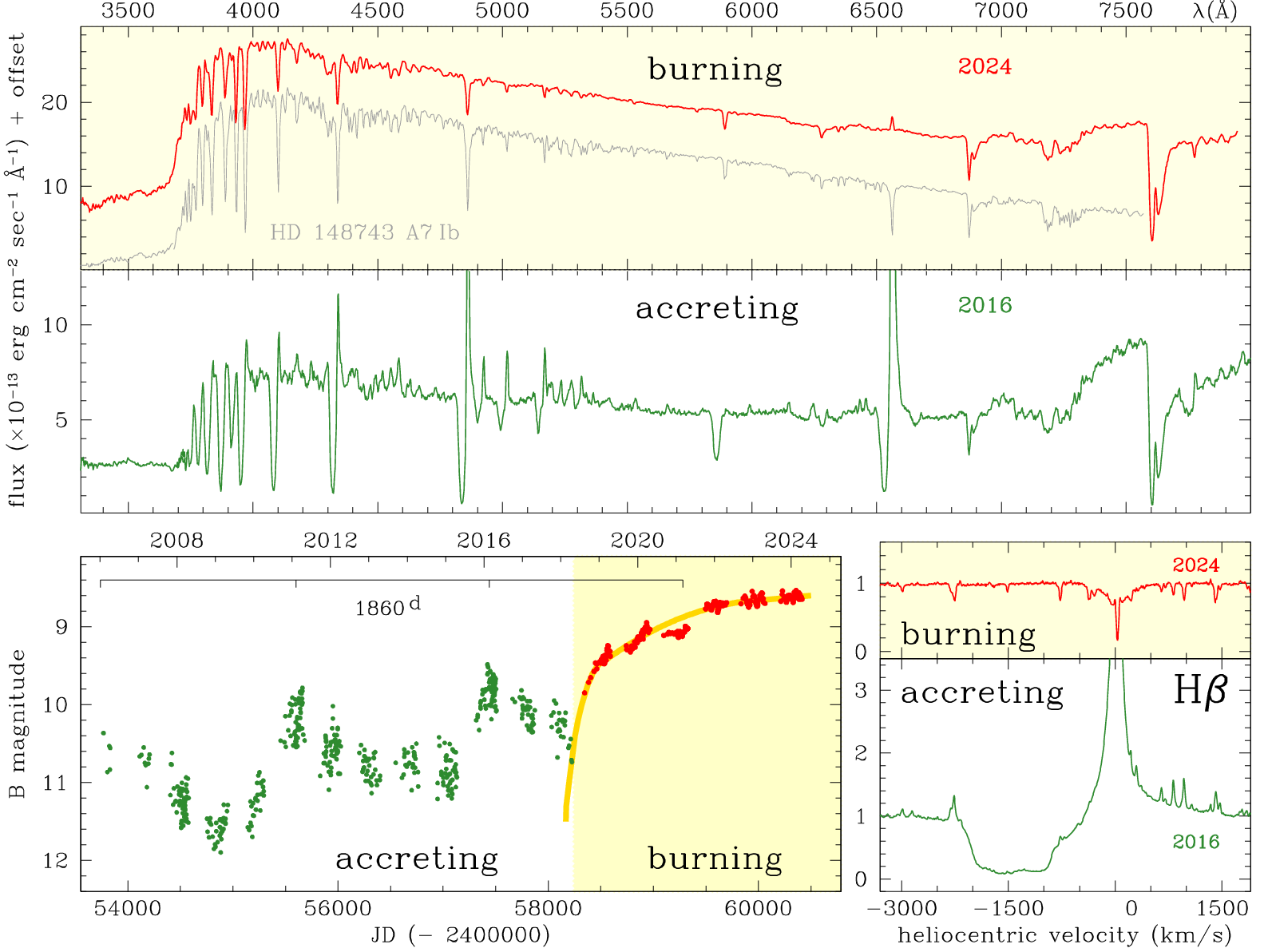}
\caption{Comparing the accreting (2016) and burning (2024) states of V694
Mon. {\em Top}: Asiago 1.22m+B\&C low-res spectra averaged over 2016 and early
2024. The latter is compared with that of the A7\,Ib supergiant HD 148743.
{\em Lower right}: high resolution H$\beta$ profiles of V694 Mon for 2016
and 2024 obtained with the Asiago 1.82m+Echelle telescope. {\em Lower left}:
ANS Collaboration $B$-band lightcurve of V694 Mon, with the orange line
highlighting the smooth rise to burning peak brightness.} 
\label{fig1}
\end{figure*}

The brightness of V694 Mon has been increasing since 2018, leveling off to a
maximum reached during early 2024, as illustrated by the $B$-band lightcurve
from ANS Collaboration presented in the lower-left panel of Fig.~\ref{fig1}. 
Contrary to 1925-2018, the post-2018 lightcurve is much sharper, confirming
the complete disappearance flickering.  The average spectrum and line
profiles of V694 Mon for early 2024 are shown in top and right panels of
Fig.~\ref{fig1}: the overall appearance is rather similar to that of an
A7\,Ib supergiant, with sharp lines, no trace of the high-velocity
absorption components, and emission limited to a weak H$\alpha$.

\section{Nuclear burning} \label{sec:burning}

I propose that the drastic changes experienced in 2018 are due to V694~Mon
transitioning from the accreting-only state (dominating since object
discovery) to non-explosive, non-degenerate, thermal-equilibrium and stable
hydrogen burning on the surface of its WD. 

The median brightness of V694~Mon for early 2024 is $V$=8.235.  At $E_{\rm
B-V}$=0.13 reddening and Gaia DR3 distance of 2.35~kpc, the absolute
magnitude of the burning WD is $M_V$=$-$4.02.  Assuming that the
bolometric correction is the same as for normal A7\,Ib stars
($B.C.$=$-$0.07), the luminosity of the burning WD is $L$=3,500~L$_\odot$
which corresponds to a white dwarf mass of $M_{\rm WD}$$\sim$0.60~M$_\odot$
following \citet{2013ApJ...777..136W}.

According to the evolutionary scenario for SySts proposed by
\citet{2019arXiv190901389M}, the longest intervals are spent accreting onto
the WD (at rates which change with time).  In the (great) majority of
SySts the accreted envelope is non-degenerate, and once the
conditions for nuclear burning are reached, the burning ignites
non-explosively and continues in thermal equilibrium as long as enough
hydrogen fuel is present in the shell, a phase lasting decades to centuries,
with the rise to optical maximum taking a few years to complete
\citep{1982ApJ...259..244I, 1982ApJ...257..752F, 1982ApJ...257..767F}.  The
shell is {\it not} ejected and remains in hydrostatic equilibrium,
progressively contracting in radius and rising in surface temperature as the
hydrogen fuel is consumed.  Once the burning shell shrinks to the radius and
temperature of the nuclei of planetary nebulae, accretion-induced ingestion
by the shell of limited amounts of mass may trigger short-lived ZAND
outbursts.
The exhaustion of the hydrogen fuel in the shell eventually stops the burning,
and the SySt resumes the long-lasting accretion phase in
preparation for a new cycle.  The alternative of the shell being instead accreted by
the WD in {\em degenerate} conditions, leads to the {\it explosive}
burning of a nova outburst, like those repeatedly experienced by the
symbiotic recurrent novae T CrB, RS Oph, V3890 Sgr, or V745 Sco.

Comparing with the closely similar spectroscopic and photometric evolution
presented by other SySts undergoing similar non-explosive, non-degenerate,
thermal-equilibrium hydrogen burning (eg. V4368 Sgr), the maximum optical
brightness and weakest emission lines may have been reached by V694 Mon in
early 2024.  If this is truly the case, V694~Mon will now enter a (very)
slow decline, characterized by a progressive increase in the photospheric
temperature of the burning shell and rise in intensity and excitation of the
emission lines and nebular continuum.

Flickering should not return for long, at least as long as the nuclear
burning will continue on the surface of the WD. As emphasized by
\citet{2003ASPC..303..202S}, flickering plays a relevant diagnostic role in
SySts: while ubiquitous among accreting-only objects, it is
completely absent among burning SySts.

\bibliography{paper.bib}{}
\bibliographystyle{aasjournal}

\end{document}